\documentclass[usegraphicx]{mn2e}

\title[Impact of $\sigma_{\mathrm{gas}}(R)$ on disc instability]
      {What drives gravitational instability in nearby star-forming spirals?
       The impact of CO and HI velocity dispersions}

\author[A. B. Romeo and K. M. Mogotsi]
       {Alessandro B. Romeo$^{1}$\thanks{E-mail: romeo@chalmers.se}
        and Keoikantse Moses Mogotsi$^{2}$\\
        $^{1}$Department of Earth and Space Sciences,
              Chalmers University of Technology,
              SE-41296 Gothenburg, Sweden\\
        $^{2}$South African Astronomical Observatory,
              P.O. Box 9, Observatory, Cape Town, South Africa}

\begin{document}

\date{Accepted 2017 April 1.
      Received 2017 March 24; in original form 2017 January 9}

\pagerange{\pageref{firstpage}--\pageref{lastpage}}

\pubyear{2017}

\maketitle

\label{firstpage}

\begin{abstract}
The velocity dispersion of cold interstellar gas, $\sigma$, is one of the
quantities that most radically affect the onset of gravitational
instabilities in galaxy discs, and the quantity that is most drastically
approximated in stability analyses.  Here we analyse the stability of a large
sample of nearby star-forming spirals treating molecular gas, atomic gas and
stars as three distinct components, and using radial profiles of
$\sigma_{\mathrm{CO}}$ and $\sigma_{\mathrm{HI}}$ derived from HERACLES and
THINGS observations.  We show that the radial variations of
$\sigma_{\mathrm{CO}}$ and $\sigma_{\mathrm{HI}}$ have a weak effect on the
local stability level of galaxy discs, which remains remarkably flat and well
above unity, but is low enough to ensure (marginal) instability against
non-axisymmetric perturbations and gas dissipation.  More importantly, the
radial variation of $\sigma_{\mathrm{CO}}$ has a strong impact on the size of
the regions over which gravitational instabilities develop, and results in a
characteristic instability scale that is one order of magnitude larger than
the Toomre length of molecular gas.  Disc instabilities are driven, in fact,
by the self-gravity of stars at kpc scales.  This is true across the entire
optical disc of every galaxy in the sample, with few exceptions.  In the
linear phase of the disc instability process, stars and molecular gas are
strongly coupled, and it is such a coupling that ultimately triggers local
gravitational collapse/fragmentation in the molecular gas.
\end{abstract}


\section{INTRODUCTION}

Gravitational instability is one of the engines behind the dynamics of disc
galaxies, where it enters a variety of processes: from the formation of stars
(Elmegreen 2012), globular clusters (Kruijssen 2014) and giant molecular
clouds (Dobbs et al.\ 2014) to the formation and evolution of spiral
structure (Bertin 2014) and bars (Athanassoula 2013; Sellwood 2014),
including the growth of bars within bars and associated structures (Shlosman
et al.\ 1989).  Today, several decades after the pioneering works of Safronov
(1960), Toomre (1964) and Goldreich \& Lynden-Bell (1965a,\,b) on local disc
instability, and the seminal papers by Lin \& Shu (1966) and Jog \& Solomon
(1984a,\,b) on the relative contributions of stars and interstellar gas, it
is widely accepted that cold gas plays an important role in the instability
scenario even though it contributes little to the self-gravity of the disc.
Numerous multi-component stability analyses have also shown that the colder
the gas, i.e.\ the lower its 1D velocity dispersion $\sigma$, the higher its
impact on the onset of disc instabilities (e.g., Bertin \& Romeo 1988, and
references therein; Elmegreen 1995; Jog 1996; Rafikov 2001; Kim \& Ostriker
2007; Elmegreen 2011; Romeo \& Falstad 2013, and references therein).

Clearly, $\sigma$ is a quantity of great importance not only for the onset of
gravitational instabilities in galaxy discs, but also for other dynamical
processes.  For example, $\sigma$ is one of the most basic diagnostics of
interstellar turbulence (see, e.g., Elmegreen \& Scalo 2004; Hennebelle \&
Falgarone 2012), which itself has an impact on \emph{both} star formation
(e.g., Krumholz \& McKee 2005; Kraljic et al.\ 2014; Salim et al.\ 2015;
Semenov et al.\ 2016) \emph{and} local disc instability (e.g., Elmegreen
1996; Romeo et al.\ 2010; Shadmehri \& Khajenabi 2012; Agertz et al.\ 2015,
and references therein).  Other examples and references are given by Mogotsi
et al.\ (2016), hereafter M16.

NGC 6946 provides an eloquent example of how radically $\sigma$ can affect
the onset of gravitational instabilities in galaxy discs, and how drastically
$\sigma$ is approximated in stability analyses (Ferguson et al.\ 1998; Romeo
\& Fathi 2015).  In particular, Ferguson et al.\ (1998) showed that if one
assumes $\sigma=6\;\mbox{km\,s}^{-1}$, the classical value motivated by
Kennicutt (1989), then this galaxy turns out to be unstable up to the edge of
the optical disc, while using a radial profile of $\sigma$ derived from
observations yields stability across the entire disc!  Martin \& Kennicutt
(2001) pointed out that radial variation in $\sigma$ remains controversial
because such measurements demand both high angular resolution and high
brightness sensitivity, requirements not met by most observations.
Fortunately, recent CO and H\,\textsc{i} galaxy surveys (BIMA SONG, HERACLES
and THINGS) have provided high-quality measurements of molecular and atomic
gas kinematics, which allow deriving reliable radial profiles of
$\sigma_{\mathrm{CO}}$ and $\sigma_{\mathrm{HI}}$ (e.g., Tamburro et
al.\ 2009; Cald\'{u}-Primo et al.\ 2013; Ianjamasimanana et al.\ 2015; Romeo
\& Fathi 2015; M16; Romeo \& Fathi 2016; Ianjamasimanana et al.\ 2017).  In
particular, Romeo \& Fathi (2015) analysed NGC 6946 in detail and showed that
the observed radial variation of $\sigma_{\mathrm{CO}}$ has indeed a
significant impact on disc instabilities.

Does the observed radial variation of $\sigma_{\mathrm{CO}}$, or that of
$\sigma_{\mathrm{HI}}$, have a significant impact on disc instabilities even
in other galaxies?  If so, how does the new instability scenario differ from
the classical one?  To explore this important aspect of the problem, we
consider a large sample of nearby star-forming spirals and use newly derived
radial profiles of $\sigma_{\mathrm{CO}}$ and $\sigma_{\mathrm{HI}}$,
together with the disc instability diagnostics developed by Romeo \& Falstad
(2013).  Such diagnostics follow from rigorous stability analyses (Romeo
1985; Bertin \& Romeo 1988; Romeo 1990, 1992, 1994), they are more general
than the effective $Q$ parameter proposed by Romeo \& Wiegert (2011) and as
easy to use.  Using such diagnostics one can measure the local stability
level of galaxy discs and the size of the regions over which gravitational
instabilities develop, and one can also predict which gas or stellar
component drives the instability process.  This has been illustrated in a
variety of applications (e.g., Genzel et al.\ 2014; Westfall et al.\ 2014;
Fathi et al.\ 2015; Romeo \& Fathi 2015; Fiacconi et al.\ 2016; Hallenbeck et
al.\ 2016; Inoue et al.\ 2016; Romeo \& Fathi 2016; Williamson et
al.\ 2016a,\,b).  In this paper, we consider not only molecular and atomic
gas but also a component that is still often disregarded when analysing the
stability of spiral galaxies: the stars!  The data and method are described
in Sect.\ 2, the results are presented in Sect.\ 3 and discussed in Sect.\ 4,
and the conclusions are drawn in Sect.\ 5.

\section{DATA AND METHOD}

We consider a sample of 12 nearby star-forming spirals that was previously
analysed by Leroy et al.\ (2008), hereafter L08, and Romeo \& Falstad (2013)
among others: NGC 628, 2841, 3184, 3198, 3351, 3521, 3627, 4736, 5055, 5194,
6946 and 7331.  These are galaxies with sensitive and spatially resolved
measurements across the entire optical disc, which L08 selected from the
following surveys: the BIMA Survey of Nearby Galaxies (BIMA SONG; Helfer et
al.\ 2003), the HERA CO-Line Extragalactic Survey (HERACLES; Leroy et
al.\ 2009), the \emph{SIRTF} / Spitzer Infrared Nearby Galaxies Survey
(SINGS; Kennicutt et al.\ 2003) and The H\,\textsc{i} Nearby Galaxy Survey
(THINGS; Walter et al.\ 2008).  We refer to L08 for a detailed description of
the data and their translation into physical quantities (see their sect.\ 3).

Following Romeo \& Falstad (2013), we treat all the molecular gas, atomic gas
and stars as three distinct components and use the same epicyclic frequency
($\kappa$), surface densities ($\Sigma_{\mathrm{CO}}$, $\Sigma_{\mathrm{HI}}$
and $\Sigma_{\star}$) and stellar radial velocity dispersion
($\sigma_{\star}$) as in L08 (see their appendices A--C and E--F).  However,
rather than using observationally motivated values of the CO and
H\,\textsc{i} 1D (line-of-sight) velocity dispersions, we use observed radial
profiles of $\sigma_{\mathrm{CO}}$ and $\sigma_{\mathrm{HI}}$, which we
describe in Sect.\ 2.1.  Note two points concerning our notation:
\begin{itemize}
\item $\Sigma_{\mathrm{CO}}$ denotes the \emph{total} surface density of
  molecular hydrogen+helium gas, as traced by CO emission (our
  $\Sigma_{\mathrm{CO}}$ = L08's $\Sigma_{\mathrm{H2}}$).
\item $\sigma_{\mathrm{CO}}$, $\sigma_{\mathrm{HI}}$ and $\sigma_{\star}$
  denote \emph{dynamically different} quantities.  To first approximation,
  molecular gas and atomic gas are collisional so their velocity dispersions
  are isotropic (see, e.g., Bertin 2014).  This is true even considering the
  effects of gas turbulence and stellar feedback (Grisdale et al.\ 2017).  In
  contrast, the stellar component is collisionless and has an anisotropic
  velocity dispersion (see again Bertin 2014).
\end{itemize}

\subsection{Radial profiles of the CO and HI velocity dispersions}

\begin{figure*}
\includegraphics[scale=1.13]{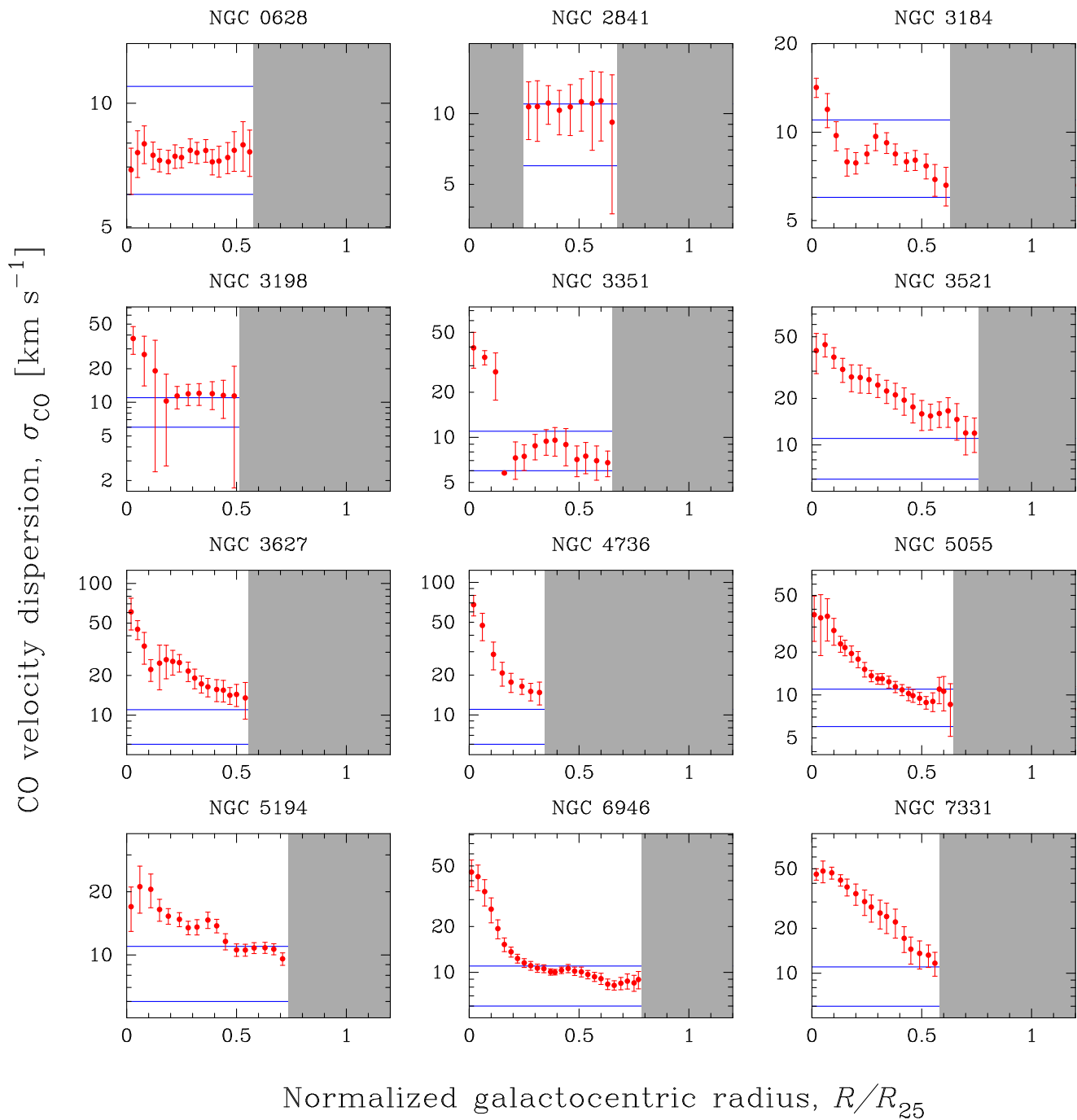}
\caption{Radial profiles of the 1D velocity dispersion of molecular gas for
  each spiral of the sample, with the galactocentric distance measured in
  units of the optical radius.  In the shaded regions, the total surface
  density of molecular gas is
  $\Sigma_{\mathrm{CO}}\leq1.0\;\mbox{M}_{\odot}\,\mbox{pc}^{-2}$ (L08).}
\end{figure*}

\begin{figure*}
\includegraphics[scale=1.13]{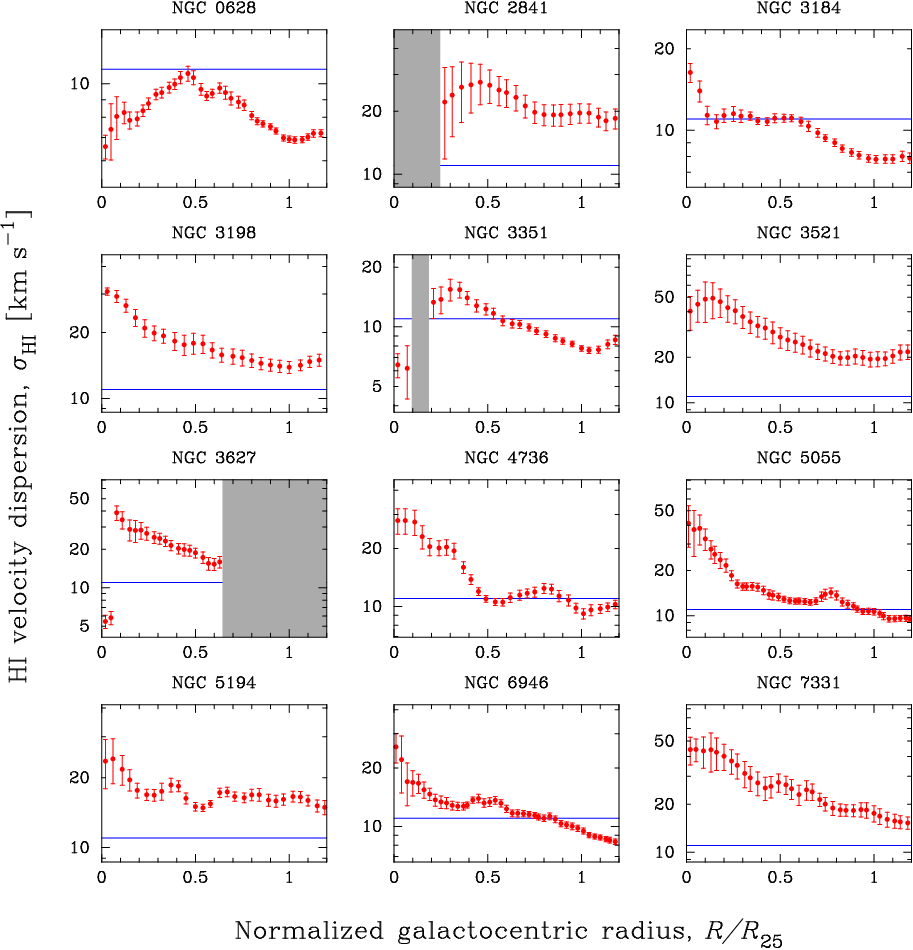}
\caption{Radial profiles of the 1D velocity dispersion of atomic gas for each
  spiral of the sample, with the galactocentric distance measured in units of
  the optical radius.  In the shaded regions, the total surface density of
  atomic gas is
  $\Sigma_{\mathrm{HI}}\leq1.0\;\mbox{M}_{\odot}\,\mbox{pc}^{-2}$ (L08).}
\end{figure*}

To derive $\sigma_{\mathrm{CO}}(R)$ and $\sigma_{\mathrm{HI}}(R)$, we use
Hanning-smoothed CO $(J=2\to1)$ data cubes from HERACLES and natural-weighted
H\,\textsc{i} data cubes from THINGS, and adapt the method used by M16 to the
present context.  In fact, a few refinements are needed to derive reliable
radial profiles of $\sigma_{\mathrm{CO}}$ and $\sigma_{\mathrm{HI}}$ for all
spirals of our sample (NGC 3521, 3627, 5194 and 7331 were not included in
M16), and to ensure that the resulting $\sigma_{\mathrm{CO}}(R)$ and
$\sigma_{\mathrm{HI}}(R)$ are fully consistent with all other radial profiles
(same sampling and range as in L08).  Our method is described step by step
below.
\begin{enumerate}
\item We smooth the H\,\textsc{i} data to 13\arcsec\ to match the spatial
  resolution of the CO data, as in M16.
\item We fit single Gaussians to the CO and H\,\textsc{i} velocity profiles,
  as in M16.
\item We then consider the CO and the H\,\textsc{i} fits \emph{separately}.
  We impose a peak amplitude cut-off equal to 4 times the root-mean-square
  level of noise, and retain only those pixels where the peak amplitude is
  greater than this cut-off value.  M16 imposed instead a more restrictive
  condition, namely that \emph{both} the CO \emph{and} the H\,\textsc{i} peak
  amplitudes should be greater than the cut-off value above.  Our way of
  processing the data is consistent with the method used by L08, who derived
  $\Sigma_{\mathrm{CO}}(R)$ and $\Sigma_{\mathrm{HI}}(R)$ \emph{independently
    of each other}.
\item We also impose a velocity dispersion cut-off equal to the typical
  velocity resolution of the data (5.2 km\,s$^{-1}$ for CO, and 2.6
  km\,s$^{-1}$ or 5.2 km\,s$^{-1}$ for H\,\textsc{i}), and retain only those
  pixels where the velocity dispersion is greater than this cut-off value.
  Our approach differs from that followed by M16 as highlighted in item
  (iii).
\item We further impose a cut-off of 2.6 km\,s$^{-1}$ on the fitted velocity
  dispersion uncertainties, and remove all pixels with uncertainties larger
  than this cut-off value.  Such a condition was not imposed by M16, but is
  useful because it reduces beam smearing and other projection effects
  significantly.  These effects are greatest in the central regions of highly
  inclined galaxies, where they cause artificial profile broadening and
  asymmetric profile shapes, especially when the velocity resolution of the
  data is low (e.g., Teuben 2002; Cald\'{u}-Primo et al.\ 2013).  Our
  condition removes most of those velocity profiles, and allows a more
  accurate determination of velocity dispersions using simple Gaussian fits.
\item We mask out further spurious emission in the CO and H\,\textsc{i} data
  using the HERACLES and THINGS moment-0 maps.
\item Finally, we compute $\sigma_{\mathrm{CO}}(R)$ and
  $\sigma_{\mathrm{HI}}(R)$ from the CO and H\,\textsc{i} velocity dispersion
  maps derived above, averaging azimuthally over 10\arcsec-wide tilted rings.
  We estimate the error bars of $\sigma_{\mathrm{CO}}(R)$ and
  $\sigma_{\mathrm{HI}}(R)$ using the traditional formula
  \begin{equation}
  \Delta X=\mbox{RMS}/\sqrt{n}\,,
  \end{equation}
  where $\Delta X$ is the uncertainty in a quantity $X$ averaged over a
  tilted ring, RMS is the root-mean-square scatter within the tilted ring,
  and $n$ is the number of resolution elements in the ring (i.e.\ the number
  of pixels in the ring where there are detections divided by the number of
  pixels per resolution element).  Here again our approach is consistent with
  that followed by L08 (M16 averaged azimuthally over 13\arcsec\ rings where
  both CO and H\,\textsc{i} have filling factors larger than 10\%).
\end{enumerate}

Figs 1 and 2 show $\sigma_{\mathrm{CO}}(R)$ and $\sigma_{\mathrm{HI}}(R)$ for
each spiral of our sample, as well as representative values of
$\sigma_{\mathrm{CO}}$ and $\sigma_{\mathrm{HI}}$ motivated/used in previous
stability analyses: $\sigma_{\mathrm{CO}}=6\;\mbox{km\,s}^{-1}$ (e.g.,
Kennicutt 1989; Wilson et al.\ 2011; Romeo \& Falstad 2013; Hallenbeck et
al.\ 2016), $\sigma_{\mathrm{CO}}=11\;\mbox{km\,s}^{-1}$ (L08), and
$\sigma_{\mathrm{HI}}=11\;\mbox{km\,s}^{-1}$ (e.g., L08; Romeo \& Falstad
2013; Hallenbeck et al.\ 2016).  Also shown, as shaded regions, are the
radial ranges where
$\Sigma_{\mathrm{CO}}\leq1.0\;\mbox{M}_{\odot}\,\mbox{pc}^{-2}$ and
$\Sigma_{\mathrm{HI}}\leq1.0\;\mbox{M}_{\odot}\,\mbox{pc}^{-2}$, i.e.\ where
the CO and H\,\textsc{i} fluxes approach the detection limit of the HERACLES
and THINGS surveys ($1.0\;\mbox{M}_{\odot}\,\mbox{pc}^{-2}$ is the working
sensitivity adopted by L08).  CO and H\,\textsc{i} data points close to the
shaded radial ranges, and H\,\textsc{i} data points close to galaxy centres,
are subject to significant systematic uncertainties.  This is true not only
for our $\sigma_{\mathrm{CO}}(R)$ and $\sigma_{\mathrm{HI}}(R)$, but also for
L08's $\Sigma_{\mathrm{CO}}(R)$ and $\Sigma_{\mathrm{HI}}(R)$.  Figs 1 and 2
illustrate that $\sigma_{\mathrm{CO}}(R)$ and $\sigma_{\mathrm{HI}}(R)$ rise
towards the centre in most of the galaxies.  Note that this is an
order-of-magnitude effect for $\sigma_{\mathrm{CO}}(R)$ in spirals like NGC
3351, 4736, 5055 and 6946!  This form of disc heating is a natural
consequence of radial inflow and is mediated by local gravitational
instabilities (e.g., Zhang 1998; Griv et al.\ 2002; Romeo et al.\ 2003, 2004;
Agertz et al.\ 2009; Forbes et al.\ 2014; Goldbaum et al.\ 2015; Romeo \&
Fathi 2015; Goldbaum et al.\ 2016; Zhang 2016).  Although there are still
open questions, the basic idea behind this process is simple, and is
beautifully illustrated in sect.\ 7.1 of Kormendy \& Kennicutt (2004).
Radial inflow increases both $\Sigma$ and $\kappa$, but $\Sigma$ `wins' and
the Toomre (1964) parameter $Q=\kappa\sigma/\pi G\Sigma$ decreases.  As $Q$
drops below a critical value of order unity, local gravitational
instabilities set in and increase $\sigma$, thus heating the disc.

\subsection{Disc instability diagnostics}

We use two disc instability diagnostics derived by Romeo \& Falstad (2013).
\begin{itemize}
\item The first diagnostic is a simple and accurate approximation for the $Q$
  stability parameter in multi-component and realistically thick discs:
  \begin{equation}
  \frac{1}{\mathcal{Q}_{N}}=\sum_{i=1}^{N}\frac{W_{i}}{T_{i}Q_{i}}\,,
  \end{equation}
  where $N$ is the number of gas and/or stellar components,
  $Q_{i}=\kappa\sigma_{i}/\pi G\Sigma_{i}$ is the Toomre parameter of
  component $i$ (remember that $\sigma$ denotes the radial velocity
  dispersion), $T_{i}$ is a factor that encapsulates the stabilizing effect
  of disc thickness for the whole range of velocity dispersion anisotropy
  ($\sigma_{z}/\sigma_{R}$) observed in galactic discs, and $W_{i}$ is a
  weight factor.  $T_{i}$ and $W_{i}$ are given by
  \begin{equation}
  T_{i}=
  \left\{\begin{array}{ll}
         {\displaystyle1+0.6\left(\frac{\sigma_{z}}{\sigma_{R}}\right)_{i}^{2}}
                         &
         \mbox{if\ }0\leq(\sigma_{z}/\sigma_{R})_{i}\leq0.5\,,
                           \\
                         & \\
         {\displaystyle0.8+0.7\left(\frac{\sigma_{z}}{\sigma_{R}}\right)_{i}}
                         &
         \mbox{if\ }0.5\leq(\sigma_{z}/\sigma_{R})_{i}\leq1\,,
         \end{array}
  \right.
  \end{equation}
  \mbox{}
  \begin{equation}
  W_{i}=\frac{2\sigma_{m}\sigma_{i}}{\sigma_{m}^{2}+\sigma_{i}^{2}}\,,
  \end{equation}
  where $m$ is the component with smallest $TQ$:
  \begin{equation}
  T_{m}Q_{m}=\min\{T_{i}Q_{i}\}\,.
  \end{equation}
\item The second diagnostic is a corresponding approximation for the
  characteristic instability scale, i.e.\ the perturbation wavelength at
  which the disc becomes locally unstable as $\mathcal{Q}_{N}$ drops below
  unity:
  \begin{equation}
  \lambda_{N}=2\pi\,\frac{\sigma_{m}}{\kappa}\,,
  \end{equation}
  where $m$ is defined by Eq.\ (5).
\end{itemize}
This set of equations tells us that the values of $\mathcal{Q}_{N}$ and
$\lambda_{N}$ are controlled by the component with smallest $TQ$.  This is
the component that drives disc instabilities: $\mathcal{Q}_{N}\sim
T_{m}Q_{m}$ ($W_{m}=1$).  All other components have less weight because their
contributions are weakened by factors $W_{i}<1$; the more $\sigma_{i}$
differs from $\sigma_{m}$, the smaller $W_{i}$.  Note that while
$\mathcal{Q}_{N}>1$ ensures stability against axisymmetric perturbations,
larger values of $\mathcal{Q}_{N}$ (\ga2--3) are required to stabilize the
disc against non-axisymmetric perturbations (e.g., Griv \& Gedalin 2012) and
gas dissipation (Elmegreen 2011).

To compute the radial profiles of $\mathcal{Q}_{3}$ and $\lambda_{3}$ ($N=3$
in our case), we need to specify $T_{\mathrm{CO}}(R)$, $T_{\mathrm{HI}}(R)$
and $T_{\star}(R)$; the radial profiles of all basic quantities have already
been specified (see Sect.\ 2).  Following Romeo \& Falstad (2013), we adopt
constant
$(\sigma_{z}/\sigma_{R})_{\mathrm{CO}}=(\sigma_{z}/\sigma_{R})_{\mathrm{HI}}
=1$, as is natural for collisional components, and a constant
$(\sigma_{z}/\sigma_{R})_{\star}=0.6$, as was assumed by L08.  Hence
$T_{\mathrm{CO}}=T_{\mathrm{HI}}=1.5$, and $T_{\star}=1.22$.

\section{RESULTS}

\begin{figure}
\includegraphics[scale=1.]{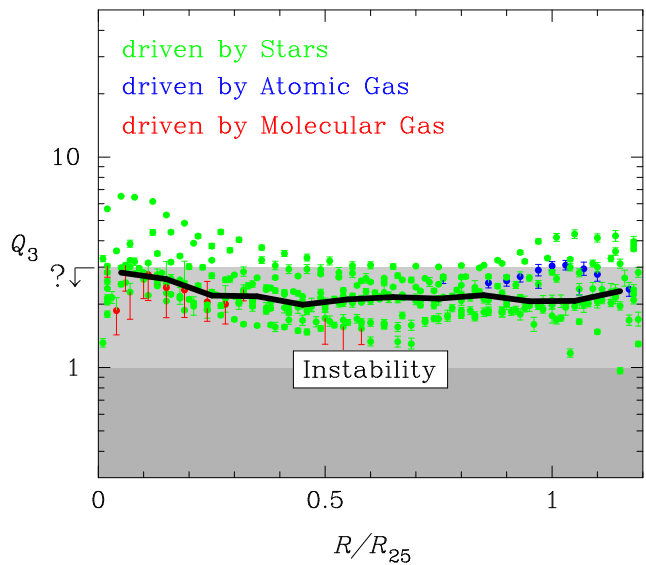}
\caption{Radial profiles of the three-component $\mathcal{Q}$ stability
  parameter for the whole sample of spirals, with the galactocentric distance
  measured in units of the optical radius.  Also shown is the local median of
  $\mathcal{Q}_{3}$.  Note that while $\mathcal{Q}_{3}>1$ ensures stability
  against axisymmetric perturbations, larger values of $\mathcal{Q}_{3}$
  (\ga2--3) are required to stabilize the disc against non-axisymmetric
  perturbations and gas dissipation.  The precise value of the critical
  stability level is still questioned (see Sect.\ 3).  The data are
  colour-coded so as to show whether disc instabilities are driven by stars,
  atomic or molecular gas.}
\end{figure}

\begin{figure*}
\includegraphics[scale=.95]{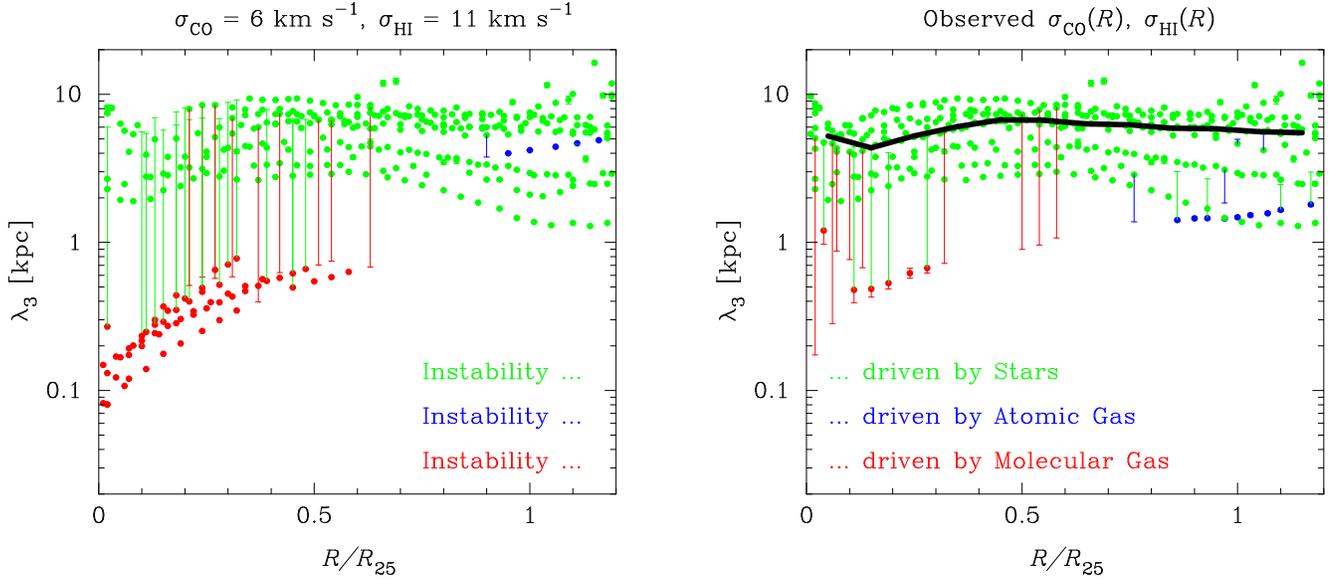}
\caption{Radial profiles of the three-component characteristic instability
  scale for the whole sample of spirals, with the galactocentric distance
  measured in units of the optical radius.  \emph{Left panel:} the result of
  using observationally motivated values of $\sigma_{\mathrm{CO}}$ and
  $\sigma_{\mathrm{HI}}$, as in Romeo \& Falstad (2013).  \emph{Right panel:}
  the result of using our observed radial profiles of $\sigma_{\mathrm{CO}}$
  and $\sigma_{\mathrm{HI}}$; also shown is the local median of
  $\lambda_{3}$.  The data are colour-coded so as to show whether disc
  instabilities are driven by stars, atomic or molecular gas.}
\end{figure*}

\begin{figure}
\includegraphics[scale=1.]{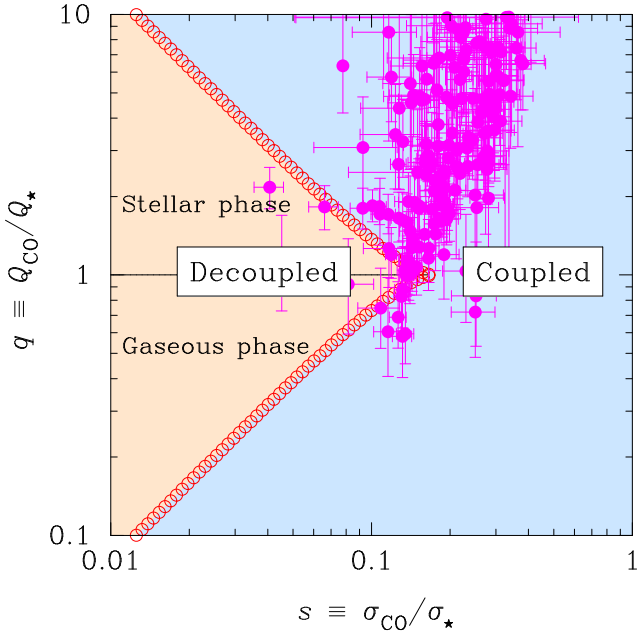}
\caption{The parameter plane of two-component disc instabilities populated by
  the galaxy data.  Here $\sigma_{\mathrm{CO}}$ and $\sigma_{\star}$ are the
  radial velocity dispersions of molecular gas and stars, $Q_{\mathrm{CO}}$
  and $Q_{\star}$ are their Toomre parameters.  Outside the `two-phase
  region', the responses of the two components to perturbations are coupled.
  So star-driven instabilities can also lead to local gravitational
  collapse/fragmentation in the molecular gas.}
\end{figure}

Fig.\ 3 shows the three-component $\mathcal{Q}$ stability parameter as a
function of galactocentric distance for our sample of spirals.  Also shown is
the local median of $\mathcal{Q}_{3}$, $\mathcal{Q}_{\mathrm{med}}(R)$,
computed by binning the data in 12 rings of width $0.1\,R/R_{25}$.  This
apparently simple plot encloses two layers of information.  The first layer
is well known: the stability level of nearby star-forming spirals is, on
average, remarkably flat and well above unity (e.g., L08; Romeo \& Falstad
2013).  In fact, the local median of $\mathcal{Q}_{3}$ varies within the
range $2\la\mathcal{Q}_{\mathrm{med}}(R)\la3$, and globally
$\mathcal{Q}_{3}=2.2\pm0.6$ (global median $\pm\ 1\sigma$ scatter).  Remember
that $\mathcal{Q}_{3}>1$ ensures stability against axisymmetric
perturbations, while larger values of $\mathcal{Q}_{3}$
($>\mathcal{Q}_{\mathrm{crit}}$) are required to stabilize the disc against
non-axisymmetric perturbations.  Unfortunately, there is still no general
consensus about the value of $\mathcal{Q}_{\mathrm{crit}}$.  For example,
Griv \& Gedalin (2012) found that the classical estimate
$\mathcal{Q}_{\mathrm{crit}}\approx2$ is an absolute upper limit on the
critical stability level.  Elmegreen (2011) showed that gas dissipation has a
similar destabilizing effect, and estimated that
$\mathcal{Q}_{\mathrm{crit}}\approx\mbox{2--3}$.  If one assumes this local
stability threshold, then nearby star-forming spirals are close to marginal
instability or unstable, given that 52\% of the data fall within the range
$2\leq\mathcal{Q}_{3}\leq3$ and that $\mathcal{Q}_{3}<2$ in 30\% of the
cases.  The second layer of information is deeper and can only be extracted
by using the Romeo-Falstad disc instability diagnostics.  It concerns the
component that drives gravitational instability, which has important
dynamical implications, as we discuss below.

Fig.\ 4 is the \emph{key plot} of our paper.  It illustrates that using
observed radial profiles of $\sigma_{\mathrm{CO}}$, rather than
observationally motivated values of $\sigma_{\mathrm{CO}}$, has a strong
impact on the inferred scale of gravitational instabilities in nearby
star-forming spirals.  Let us first see what a
$\sigma_{\mathrm{CO}}=6\;\mbox{km\,s}^{-1}$ analysis predicts.  Remember that
this is the value of $\sigma_{\mathrm{CO}}$ motivated by Kennicutt (1989),
Martin \& Kennicutt (2001) and Wilson et al.\ (2011) among others, and the
one used by Romeo \& Falstad (2013).  The left panel of Fig.\ 4 shows the
three-component characteristic instability scale as a function of
galactocentric distance for our sample of spirals.  The colour coding tells
us which component drives gravitational instability: molecular gas, atomic
gas or stars.  Note (i) that disc instabilities are driven by the stars in
82\% of the cases, but molecular gas is nevertheless an important driver of
gravitational instability in the inner disc; (ii) that the distribution of
$\lambda_{3}(R)$ is bimodal for $R\la0.5\,R_{25}$; and (iii) that there is an
order-of-magnitude gap in $\lambda_{3}$ between regimes driven by the
molecular gas (a few 100 pc) and regimes driven by the stars (a few kpc).
What does our analysis predict instead?  See the right panel of Fig.\ 4.  The
rise of $\sigma_{\mathrm{CO}}(R)$ towards the centre results in a paradigm
shift.  Inner disc instabilities are now entirely driven by the stars at kpc
scales ($\lambda_{3}=5.9\pm2.3\,\mbox{kpc}$), with two notable exceptions:
the inner discs of NGC 5194 and NGC 6946 (2\% of the data).  These are the
sample galaxies with highest surface density of molecular gas averaged over
the inner disc ($\langle\Sigma_{\mathrm{CO}}\rangle\approx
150\;\mbox{M}_{\odot}\,\mbox{pc}^{-2}$ for $R\leq0.3\,R_{25}$), and highest
star formation rate ($\mbox{SFR}\ga3\;\mbox{M}_{\odot}\,\mbox{yr}^{-1}$; see
table 4 of L08, or table 1 of Walter et al.\ 2008).  These are also active
galaxies, where molecular gas plays a key role as fuel in the activity
process (e.g., Krips et al.\ 2008).  Fig.\ 4 also shows that atomic gas plays
a negligible role as driver of gravitational instability up to the edge of
the optical disc, regardless of whether one uses
$\sigma_{\mathrm{HI}}=11\;\mbox{km\,s}^{-1}$ (L08; Romeo \& Falstad 2013) or
our observed $\sigma_{\mathrm{HI}}(R)$.

Can star-driven instabilities lead to local gravitational
collapse/fragmentation in the molecular gas?  To answer this question, we
should understand in more detail how molecular gas and stars contribute to
disc instabilities.  This important piece of information is illustrated in
Fig.\ 5.  Consider a two-component disc of molecular gas and stars, and
perturb it with axisymmetric waves of frequency $\omega$ and wavenumber $k$.
The response of the disc is basically described by the Jog-Solomon (1984a)
dispersion relation, $\omega^{2}(k)$.  Inside the `two-phase region' shown in
Fig.\ 5, $\omega^{2}(k)$ has two distinct minima (Bertin \& Romeo 1988; Romeo
\& Wiegert 2011).  In the `gaseous phase', the minimum at short wavelengths
($\lambda=2\pi/k$) is lower than the other one and molecular gas will drive
the onset of gravitational instability.  Vice versa, in the `stellar phase',
the long-wavelength minimum is lower and stars will drive instability.  The
shape and size of this region are only moderately affected by disc thickness
(Romeo \& Wiegert 2011), gas turbulence (Hoffmann \& Romeo 2012) or the fact
that the stellar component is collisionless (Romeo \& Falstad 2013).  In the
rest of the parameter plane, $\omega^{2}(k)$ has a single minimum, where the
dynamical responses of the two components are strongly coupled.  This means
that any instability driven by one of the components will also perturb and
destabilize the other.  In particular, star-driven instabilities will lead to
local gravitational collapse/fragmentation in the molecular gas.  This is
clearly the case for almost the entire galaxy sample, as almost the entire
data set falls outside the two-phase region.  The inner discs of NGC 5194 and
NGC 6946 are again exceptions to the general rule.  But instabilities in such
discs are driven by the molecular gas (see discussion of Fig.\ 4), so they
will naturally lead to its collapse or fragmentation.

\section{DISCUSSION}

\begin{figure}
\includegraphics[angle=-90.,scale=.95]{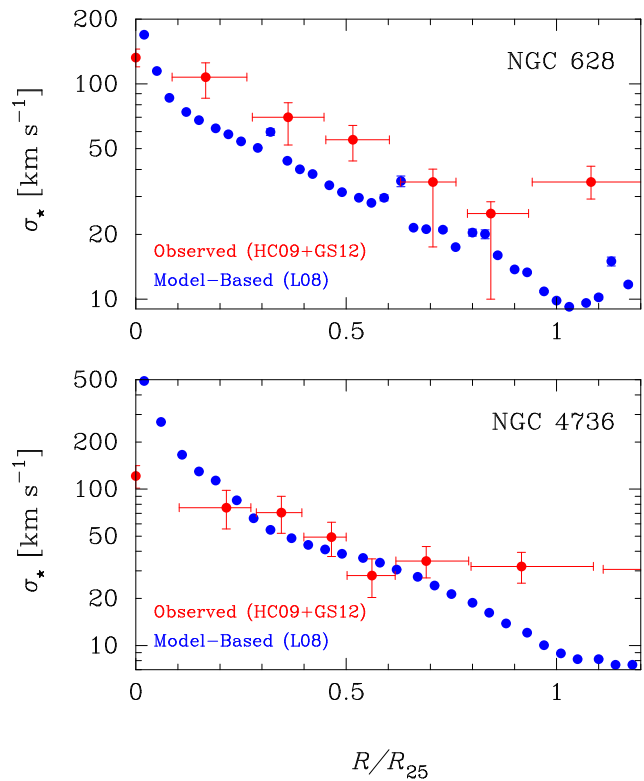}
\caption{Observed vs.\ model-based radial profiles of the stellar radial
  velocity dispersion for two spirals of the sample, with the galactocentric
  distance measured in units of the optical radius.}
\end{figure}

Are NGC 5194 and NGC 6946 true outliers?  The condition that disc
instabilities are driven by stars, rather than molecular gas, is
\begin{equation}
T_{\star}Q_{\star}<T_{\mathrm{CO}}\,Q_{\mathrm{CO}}
\;\;\;\Longrightarrow\;\;\;
T_{\star}\sigma_{\star}/\Sigma_{\star}<
T_{\mathrm{CO}}\,\sigma_{\mathrm{CO}}/\Sigma_{\mathrm{CO}}
\end{equation}
(see Sect.\ 2.2 and remember that $\Sigma_{\mathrm{CO}}$ denotes the total
surface density of molecular gas, helium included).  As pointed out in
Sect.\ 3, this condition is not fulfilled by the inner discs of NGC 5194 and
NGC 6946.  On the other hand, it turns out that if
$T_{\star}\sigma_{\star}/\Sigma_{\star}$ were 29\% smaller, or if
$T_{\mathrm{CO}}\,\sigma_{\mathrm{CO}}/\Sigma_{\mathrm{CO}}$ were 41\% larger
($1/0.71=1.41$), then gravitational instabilities in such discs would be
driven by the stars, as in the rest of the galaxy sample.  Can systematic
uncertainties account for that?  Let us discuss this point below.

It is well known that the surface densities of molecular gas and stars are
subject to significant systematic uncertainties via the adopted
CO-to-$\mathrm{H}_{2}$ conversion factor, $X_{\mathrm{CO}}$, and stellar
mass-to-light ratio, $\Upsilon_{\star}$ (see, e.g., Binney \& Merrifield
1998).  L08 adopted $X_{\mathrm{CO}}=
2\times10^{20}\;\mbox{cm}^{-2}\;(\mbox{K\,km\,s}^{-1})^{-1}$.  This is the
standard value recommended by Bolatto et al.\ (2013) for the discs of normal
solar-metallicity galaxies, and has an uncertainty of $\pm$\,0.3\,dex (a
factor of 2).  Sandstrom et al.\ (2013) carried out one of the most
comprehensive extragalactic study of $X_{\mathrm{CO}}$ to date, and found an
average value of $X_{\mathrm{CO}}$ that is 30\% smaller than the standard
one: $\langle X_{\mathrm{CO}}\rangle=
1.4\times10^{20}\;\mbox{cm}^{-2}\;(\mbox{K\,km\,s}^{-1})^{-1}$, again with an
uncertainty of $\pm$\,0.3\,dex.  This value of $X_{\mathrm{CO}}$ is small
enough to move the inner discs of NGC 5194 and NGC 6946 into star-driven
instability regimes!  Concerning the stellar mass-to-light ratio, L08 adopted
$\Upsilon_{\star}^{K}=0.5\;\mbox{M}_{\odot}/\mbox{L}_{\odot,K}$.  This is
near the mean $K$-band $M/L$ ratio expected for these galaxies (Bell et
al.\ 2003), and has an uncertainty of $\pm$\,0.1--0.2\,dex (a factor of
1.3--1.6).  A value of
$\Upsilon_{\star}^{K}=0.7\;\mbox{M}_{\odot}/\mbox{L}_{\odot,K}$ would be
large enough to `normalize' NGC 5194 and NGC 6946, and still be within the
uncertainty range.

Another source of significant systematic uncertainty is the stellar radial
velocity dispersion.  In fact, contrary to our $\sigma_{\mathrm{CO}}(R)$, the
radial profiles of $\sigma_{\star}$ derived by L08 are not based on
observations, but on a simple model that relates $\sigma_{\star}$ to the
surface density and scale length of the stellar disc (see appendix B.3 of
L08).  To the best of our knowledge, stellar velocity dispersions have only
been measured in three galaxies of the sample: NGC 628 (Ganda et al.\ 2006;
Herrmann \& Ciardullo 2009), NGC 3198 (Bottema 1988, 1993) and NGC 4736
(Herrmann \& Ciardullo 2009).  To estimate the accuracy of L08's model, we
consider the radial profiles of $\sigma_{z\star}$ derived by Herrmann \&
Ciardullo (2009) for NGC 628 and NGC 4736, and convert $\sigma_{z\star}$ into
$\sigma_{\star}$ using the best-fitting model of Gerssen \& Shapiro Griffin
(2012), which relates $(\sigma_{z}/\sigma_{R})_{\star}$ to galaxy type (see
their fig.\ 4).  Fig.\ 6 shows that L08's model is accurate to within a
factor of 2 except in the innermost/outermost regions of the stellar disc,
where this model can overestimate/underestimate the observed
$\sigma_{\star}(R)$ by a larger factor.  Our $\sigma_{\mathrm{CO}}(R)$ is
less uncertain.  The main source of systematic uncertainty is beam smearing,
which is greatest in the central regions of highly inclined galaxies (e.g.,
Teuben 2002; Cald\'{u}-Primo et al.\ 2013).  However, beam smearing can
increase the velocity dispersion by at most a factor of 1.2 at
$R=0.2\,R_{25}$ for galaxies with 30\degr\ inclination, 1.5 for 60\degr\ and
1.8 for 80\degr, with these factors decreasing quickly towards unity at
larger radii (Cald\'{u}-Primo et al.\ 2013).

The bottom line is that systematic uncertainties in $\Sigma_{\mathrm{CO}}$ or
in $\Sigma_{\star}$ and $\sigma_{\star}$ can account for a significant
increase in $T_{\mathrm{CO}}\,Q_{\mathrm{CO}}$ or decrease in
$T_{\star}Q_{\star}$, and thus move the inner discs of NGC 5194 and NGC 6946
into star-driven instability regimes.  This is not surprising.  Stars are the
primary driver of gravitational instabilities even in the inner disc of NGC
1068, a powerful nearby Seyfert+starburst galaxy (Romeo \& Fathi 2016).

Finally, the result that molecular gas plays a secondary role in disc
instabilities (at low redshift) is based on the spatial resolution of current
extragalactic surveys: BIMA SONG, HERACLES, SINGS and THINGS.  At scales
smaller than a few 100 pc, interstellar turbulence excites two non-classical
instability regimes where molecular gas plays a primary role.  Such regimes
stretch from galactic-scale Toomre instability to clump-scale (3D) Jeans
instability, and may be one of the missing links between disc instabilities
and star formation.  Such regimes are illustrated in fig.\ 1 of Romeo et
al.\ (2010): `\emph{the stability map of turbulence}' (see also Hoffmann \&
Romeo 2012; Romeo \& Agertz 2014; Agertz et al.\ 2015).

\section{CONCLUSIONS}

In this paper, we have analysed the stability of a sample of 12 nearby
star-forming spirals using the Romeo-Falstad $Q$ stability parameter,
$\mathcal{Q}_{N}$, and characteristic instability scale, $\lambda_{N}$, for
$N=3$ disc components: molecular gas, atomic gas and stars.  The most novel
feature of our analysis is that we have made use of observed radial profiles
of the CO and H\,\textsc{i} velocity dispersions, rather than observationally
motivated values of $\sigma_{\mathrm{CO}}$ and $\sigma_{\mathrm{HI}}$.  Our
major conclusions are pointed out below.
\begin{itemize}
\item The CO velocity dispersion has a strong impact on the disc instability
  scenario, as illustrated in Fig.\ 4.  Using the classical value
  $\sigma_{\mathrm{CO}}=6\;\mbox{km\,s}^{-1}$ motivated by Kennicutt (1989),
  one would infer that molecular gas plays a significant role in disc
  instabilities even at distances as large as half the optical radius.  In
  particular, the characteristic instability scale would have a bimodal
  radial distribution with an order-of-magnitude gap between regimes driven
  by the molecular gas ($\lambda_{3}\approx\mbox{80--800\,pc}$) and regimes
  driven by the stars ($\lambda_{3}\approx\mbox{2--10\,kpc}$).  Using instead
  our radial profiles of $\sigma_{\mathrm{CO}}$ results in disc instabilities
  that are almost entirely driven by the stars.  The characteristic
  instability scale has median value $\lambda_{3}=5.9\,\mbox{kpc}$ and
  $1\sigma$ scatter $\Delta\log\lambda_{3}=0.16\,\mbox{dex}$ (a factor of
  1.4); 2\% of the data are `outliers', which systematic uncertainties can
  move into star-driven regimes.
\item In contrast to the characteristic instability scale, the $Q$ stability
  parameter is robust against radial variations in both
  $\sigma_{\mathrm{HI}}$ and $\sigma_{\mathrm{CO}}$.  On average, its radial
  profile remains remarkably flat and well above unity, but near or below the
  approximate threshold for stability against local, non-axisymmetric, linear
  perturbations and gas dissipation ($\mathcal{Q}_{3}\approx\mbox{2--3}$).
  Specifically, the $Q$ stability parameter has median value
  $\mathcal{Q}_{3}=2.2$ and $1\sigma$ scatter
  $\Delta\log\mathcal{Q}_{3}=0.11\,\mbox{dex}$ (a factor of 1.3); 52\% of the
  data fall within the range $2\leq\mathcal{Q}_{3}\leq3$, and
  $\mathcal{Q}_{3}<2$ in 30\% of the cases.
\item The conclusion that stars are the primary driver of disc instabilities
  in nearby star-forming spirals requires two further clarifications.  First,
  in the linear phase of the disc instability process stars are strongly
  coupled to molecular gas, as shown in Fig.\ 5.  This means that any
  instability driven by the stars will also perturb and destabilize molecular
  gas, and thus lead to local gravitational collapse/fragmentation.  Second,
  the fact that stars set the initial conditions for gravitational
  instability is true at the spatial resolution of current extragalactic
  surveys: BIMA SONG, HERACLES, SINGS and THINGS.  At scales smaller than
  $\sim100\,\mbox{pc}$, interstellar turbulence opens new instability
  channels in which molecular gas plays a primary role (e.g., Romeo et
  al.\ 2010).
\end{itemize}

\section*{ACKNOWLEDGEMENTS}

This work made use of data from the following surveys: BIMA SONG, `The BIMA
Survey of Nearby Galaxies' (Helfer et al.\ 2003); HERACLES, `The HERA CO-Line
Extragalactic Survey' (Leroy et al.\ 2009); SINGS, `The \emph{SIRTF} /
Spitzer Infrared Nearby Galaxies Survey' (Kennicutt et al.\ 2003); and
THINGS, `The H\,\textsc{i} Nearby Galaxy Survey' (Walter et al.\ 2008).  This
work also made use of radial profiles derived by Leroy et al.\ (2008) and
Herrmann \& Ciardullo (2009).  We are very grateful to Oscar Agertz, Kim
Herrmann and Andreas Schruba for useful discussions.  We are also grateful to
an anonymous referee for constructive comments and suggestions, and for
encouraging future work on the topic.  KMM gratefully acknowledges support
from the Square Kilometre Array South Africa (SKA-SA).

\bsp

\label{lastpage}

\end{document}